\documentclass[epj]{svjour}

\usepackage{epsf}

\begin{document}

\title{Finite-size scaling above the upper critical dimension revisited: The
case of the five-dimensional Ising model}
%\subtitle{}
\author{Erik Luijten\inst{1,2,}\thanks{e-mail: erik.luijten@uni-mainz.de}
        \and Kurt Binder\inst{2}
        \and Henk W.J. Bl\"ote\inst{3}}

\institute{Max-Planck-Institut f\"ur Polymerforschung, Postfach 3148,
           D-55021 Mainz, Germany
           \and
           Institut f\"ur Physik, WA 331, Johannes Gutenberg-Universit\"at,
           D-55099 Mainz, Germany
           \and
           Department of Physics, Delft University of Technology, P.O. Box
           5046, 2600 GA Delft, The Netherlands}
\date{\today}

\authorrunning{Erik Luijten \emph{et al.}}
\titlerunning{Finite-size scaling above the upper critical dimension revisited}

\abstract{%
Monte Carlo results for the moments $\langle M^k\rangle$ of the magnetization
distribution of the nearest-neighbor Ising ferromagnet in a $L^d$ geometry,
where $L$ ($4 \leq L \leq 22$) is the linear dimension of a hypercubic lattice
with periodic boundary conditions in $d=5$ dimensions, are analyzed in the
critical region and compared to a recent theory of Chen and Dohm (CD) [X.S.
Chen and V. Dohm, Int.\ J. Mod.\ Phys.\ C (1998)]. We show that this
finite-size scaling theory (formulated in terms of two scaling variables) can
account for the longstanding discrepancies between Monte Carlo results and the
so-called ``lowest-mode'' theory, which uses a single scaling variable
$tL^{d/2}$ where $t=T/T_{\rm c}-1$ is the temperature distance from the
critical temperature, only to a very limited extent. While the CD theory gives
a somewhat improved description of corrections to the ``lowest-mode'' results
(to which the CD theory can easily be reduced in the limit $t \to 0$, $L \to
\infty$, $tL^{d/2}$ fixed) for the fourth-order cumulant, discrepancies are
found for the susceptibility ($L^d \langle M^2 \rangle$).  Reasons for these
problems are briefly discussed.
}

\maketitle

\section{Introduction}
Since about 15 years the five-dimensional Ising model is used as a
``laboratory''~\cite{bnpy,binder5d,rickwardt,mon,parisi,5dcomm,5drepl,univ5d}
to test theoretical concepts about critical phenomena, in particular the
concept of finite-size scaling~\cite{fisher-fss,fisher_barber,binder72,%
binder-cum,brezin,privfish,botet,barber,bzj,rudnick,rudnick2,huhn,binder_wang,%
privman_book,binder_ft,dohm93,mon_binder,dohm,renorm,medran,cd_physica,%
cd_epjb,cd_ijmpc,univfss}, which has become an extremely valuable and
indispensable tool for the study of phase transitions in condensed
matter~\cite{privman_book,binder_ciccotti,binder_rpp} and gauge theories of
elementary particle physics~\cite{gausterer,meyer}. In this context, the $d=5$
Ising model is of particular interest, since it exceeds the upper critical
dimension, $d^*=4$, and hence the Landau mean-field exponents exactly describe
the critical behavior~\cite{fisher_rmp,qft}. Also correction-to-scaling
exponents~\cite{wegner_riedel} are known precisely~\cite{qft} and fluctuations
around the leading mean-field description can be dealt with by simple
perturbation theory; a renormalization-group treatment of fluctuation effects
is not required here~\cite{fisher_rmp,qft}. While for $d>d^*$ the hyperscaling
relation $d\nu=\gamma+2\beta$ ($\nu$, $\gamma$, $\beta$ being the standard
critical exponents for correlation length $\xi$, susceptibility $\chi$ and
order parameter $M$, respectively) does not hold and hence finite-size scaling
in its standard version (``the linear dimension $L$ scales with
$\xi$''~\cite{fisher-fss,fisher_barber,binder72,binder-cum,barber,huhn,%
binder_ft}) does not hold either~\cite{binder-cum,brezin,privfish}, a simple
extension was proposed~\cite{bnpy,binder5d,bzj,rudnick} which can be phrased
as~\cite{binder5d} ``the linear dimension~$L$ scales asymptotically with a
thermodynamic length $\ell_{\rm T}=(M^2_{\rm b}/\chi_{\rm b})^{-1/d} \propto
t^{-2/d}$''. Moreover it was suggested that ratios of moments of the order
parameter distribution, such as $Q \equiv \langle M^2\rangle^2/\langle
M^4\rangle$ or $\langle |M|\rangle^2/\langle M^2\rangle$ can easily be
calculated from the so-called lowest-mode approximation~\cite{bzj}, which was
believed to be exact for the limit $L \to \infty$, $t \to 0$, $L/\ell_{\rm T}$
fixed and should yield, apart from scale factors, universal finite-size scaling
functions of $L/\ell_{\rm T}$.

In view of these rather definite predictions~\cite{bzj}, apparent discrepancies
between the theoretical results and the Monte Carlo
simulations~\cite{bnpy,binder5d,rickwardt} have been disturbing and it has been
a matter of debate
\cite{renorm,mon,parisi,5dcomm,5drepl,univ5d,cd_physica,cd_epjb,cd_ijmpc}
whether the discrepancies reflect corrections to finite-size scaling.  In
Ref.~\cite{univ5d} it was shown that the Monte Carlo data for~$Q$ are indeed
compatible with the predictions of Br\'ezin and Zinn-Justin~\cite{bzj} if one
takes into account two, theoretically predicted, corrections to scaling.
However, this still left the very slow convergence of~$Q$ as a function of~$L$
toward its predicted asymptotic value as a remarkable feature (in
Ref.~\cite{rickwardt} the data for~$Q$ for the considered range of system sizes
appeared to have a common intersection point unequal to this value).  More
importantly, the correctness of the treatment in Ref.~\cite{bzj} has recently
been doubted~\cite{cd_ijmpc} (see below).  This controversy is cumbersome
because the fact that it is apparently very difficult to disentangle the
leading and subleading terms in finite-size scaling analyses even in a case
where all involved critical exponents are known precisely naturally leads to
some doubt on analyses where one wants to extract unknown critical exponents
from finite-size scaling \cite{privman_book,binder_ft,dohm,binder_rpp}. In
addition, the problem also is of interest in the context of physical systems
that are nearly described by Landau theory, such as systems with a long but
finite range of interaction~\cite{mon_binder,medran}, polymer mixtures near
their critical point of unmixing~\cite{deutsch}, etc.

New light has been shed on this state of affairs by Chen and Dohm
(CD)~\cite{cd_physica,cd_epjb,cd_ijmpc}, who presented detailed arguments that
for $d>d^*$ the standard treatment of the $\phi^4$ field theory in continuous
space~\cite{bzj,rudnick,qft} yields a misleading description of finite-size
behavior, different from the finite-size behavior of a $\phi^4$ model on a
lattice, which one believes to be equivalent to an Ising
model~\cite{fisher_rmp,qft}. Chen and Dohm emphasized that therefore the
justification given for the lowest-mode theory is invalid, and stated that the
moment ratios mentioned above ``do not have the universal properties predicted
previously and that recent analyses of Monte Carlo results for the
five-dimensional Ising model are not conclusive''~\cite{cd_ijmpc}.

In view of this criticism, a reanalysis of the available Monte Carlo results
(including also some recent unpublished results~\cite{5dpaper} used
in~\cite{thesis}) is clearly warranted. Such an analysis, where we compare the
Monte Carlo data in detail with the result of the CD theory (which treats order
parameter fluctuations perturbatively to one-loop order for the Ising
case~\cite{cd_ijmpc}) is given here. For the sake of a coherent presentation,
we summarize the pertinent theoretical results in Sec.~\ref{sec:theory}, while
Sec.~\ref{sec:compare} gives a detailed comparison of the results for $Q$ and
for the susceptibility $\chi = L^d\langle M^2\rangle$ with the CD
theory. Section~\ref{sec:concl} summarizes our conclusions.

\section{Theoretical background}
\label{sec:theory}
The singular part of the free-energy density $f_L$ of a system with linear
size~$L$ in an external field $h$ is written as (see, e.g.,
Ref.~\cite{privfish})
\begin{equation}
 f_L = L^{-d}f\left(tL^{y_{\rm t}},hL^{y_{\rm h}},uL^{y_{\rm i}}\right) \;,
 \quad L \to \infty \;,
\label{eq:free_energy}
\end{equation}
where $t=T/T_{\rm c}-1$ and $u$ is an irrelevant variable, in the
renormalization-group sense (exponents $y_{\rm t} > 0$, $y_{\rm h} > 0$,
$y_{\rm i} < 0$). Now for $d>d^*=4$ we have $y_{\rm t}=2$, $y_{\rm h}=(d+2)/2$
and $y_{\rm i}=4-d$, but $u$ is a ``dangerous irrelevant variable'' (see, e.g,
Refs.~\cite{kvlh,fisher_review}), which means that the scaling function
$f(x_1,x_2,x_3)$ is singular in the limit $x_3 \to 0$ and cannot simply be
replaced by $f(x_1,x_2,0)$. In terms of the bulk correlation length $\xi_{\rm
b} = \xi_0 t^{-\nu}$ (above~$T_{\rm c}$ in zero field), the first argument of
Eq.~(\ref{eq:free_energy}) can be interpreted as $(L/\xi_{\rm b})^2$.  Taking
suitable derivatives of Eq.~(\ref{eq:free_energy}) with respect to the field we
can thus write for the order parameter, the susceptibility and the ratio~$Q$
(in zero field)
\begin{eqnarray}
\label{eq:scale-m}
\langle |M| \rangle &=& 
   L^{-(d-2)/2} P_M \left\{ t(L/\xi_0)^2,(L/\ell_0)^{4-d}\right\} \;, \\
\label{eq:scale-chi}
\chi &=& \left(\frac{\partial^2 f_L}{\partial h^2}\right) 
      = L^d \langle M^2 \rangle \nonumber \\
     &=& L^2 P_{\chi} \left\{ t(L/\xi_0)^2,(L/\ell_0)^{4-d}\right\}
\end{eqnarray}
and
\begin{equation}
\label{eq:scale-q}
Q     = \frac{\langle M^2 \rangle^2}{\langle M^4 \rangle}
      = P_{Q} \left\{ t(L/\xi_0)^2,(L/\ell_0)^{4-d}\right\} \;,
\end{equation}
where $P_M$, $P_\chi$ and $P_Q$ are the (universal~\cite{privman_book})
finite-size scaling functions of the quantities $\langle |M|\rangle$, $\chi$
and $Q$.  The correlation-length amplitude $\xi_0$ of the bulk correlation
length now appears as a scale factor for the variable~$x_1$ and for the
variable~$x_3$ we have introduced the correlation-length amplitude $\ell_0$ of
the bulk correlation length at $T_{\rm c}$ in a small field~\cite{cd_ijmpc} as
a scale factor. In this way, the arguments of the scaling functions $P_M$,
$P_\chi$ and $P_Q$ are dimensionless, as they should be. Note that $u \propto
\ell_0^{d-4}$.

For large $L$ the variable $x_3 \propto (L/\ell_0)^{4-d}$ clearly becomes very
small, and hence it is an obvious question to ask how all these functions
$f_L$, $P_M$, $P_\chi$, $P_Q$ behave in the limit $x_3 \to 0$. It was assumed
in Ref.~\cite{bnpy} that then the dangerous irrelevant variable $x_3$ enters in
the form of multiplicative singular powers of $x_3$, e.g.,
\begin{equation}
 f_L(x_1,x_2,x_3) = x_3^{p_3} \tilde{f}_L(x_1 x_3^{p_1}, x_2 x_3^{p_2}) \;.
\end{equation}
This assumption was in the first place motivated by the fact that this is the
mechanism that operates for the scaling in the bulk for
$d>4$~\cite{fisher_review}, and secondly by various phenomenological
arguments. In particular, it was argued that standard thermodynamic fluctuation
theory requires for $T<T_{\rm c}$ and sufficiently large $L$ that the
distribution function $P_L(M)$ of the magnetization per spin for $M$ near the
spontaneous magnetization $\pm M_{\rm b}$ is a sum of two
Gaussians~\cite{bnpy,binder5d,binder-cum},
\begin{eqnarray}
 P_L(M) &=& \frac{L^{d/2}}{2\sqrt{2\pi \chi_{\rm b}}}
 \left\{ \exp\left[-(M-M_{\rm b})^2 L^d/2\chi_{\rm b}\right] \right. 
 \nonumber \\
 && + \left.  \exp\left[-(M+M_{\rm b})^2 L^d/2\chi_{\rm b}\right] \right\} \;.
\end{eqnarray}
Using $M_{\rm b}=\hat{M}_{\rm b} |t|^\beta=\hat{M}_{\rm b} (-t)^{1/2}$ and
$\chi_{\rm b}=\hat{\chi}'_{\rm b} |t|^{-\gamma} = \hat{\chi}'_{\rm b}
(-t)^{-1}$ the arguments of the exponential functions have the form
\begin{equation}
 \frac{1}{2} \left[(M/\hat{M}_{\rm b})|t|^{-1/2} \mp 1\right]^2
 (L/\ell_{\rm T})^d \;,
\end{equation}
with $\ell_{\rm T}^d=M_{\rm b}^{-2}\chi_{\rm b}=\hat{M}_{\rm b}^{-2}
\hat{\chi}'_{\rm b} t^{-2}$.
Taking moments of this distribution one hence expects that the scaling
functions in Eqs.\ (\ref{eq:scale-m})--(\ref{eq:scale-q}) reduce to
scaling functions of a single variable $(L/\ell_{\rm T})^{d/2} \propto
tL^{d/2}$ or, keeping the amplitudes $\xi_0$ and $\ell_0$ explicitly present,
$tL^{d/2}\xi_0^{-2}\ell_0^{(4-d)/2}$, i.e.,
\begin{eqnarray}
\label{eq:scale2-m}
 \langle |M| \rangle &=& 
 L^{-d/4} \tilde{P}_M \left(tL^{d/2}\xi_0^{-2}\ell_0^{(4-d)/2} \right) \;, \\
\label{eq:scale2-chi}
 \chi &=& 
 L^{d/2} \tilde{P}_\chi \left(tL^{d/2}\xi_0^{-2}\ell_0^{(4-d)/2} \right)
\end{eqnarray}
and
\begin{equation}
\label{eq:scale2-q}
 Q = \tilde{P}_Q \left(tL^{d/2}\xi_0^{-2}\ell_0^{(4-d)/2} \right) \;.
\end{equation}
Note that scale factors for the magnetization and the susceptibility have been
absorbed in $P_M$ (or $\tilde{P}_M$) and $P_\chi$ (or $\tilde{P}_\chi$),
respectively, while in ratios such as $Q$ (and hence in $P_Q$ and
$\tilde{P}_Q$) such scale factors are divided out and fully universal functions
remain.

These arguments as they were presented in Refs.~\cite{bnpy,binder5d} did not
tell anything about the explicit form of the scaling functions
$\tilde{P}_M(x)$, $\tilde{P}_\chi(x)$ and $\tilde{P}_Q(x)$, however, and hence
no prediction for the supposedly universal constant $\tilde{P}_Q(0)$ was
made. In fact, making the (premature!) assumption that linear dimensions
$L=3$~to~$7$ lattice spacings are already large enough to obtain the limit $x_3
\propto (L/\ell_0)^{-1} \to 0$ in $d=5$ dimensions, it was argued that at
$T_{\rm c}$ there occurs a distribution of the scaled order parameter
$P_L(\phi) \propto \exp(A\phi^2-\phi^4$), which implies a shift of $T_{\rm c}$
as $T_{\rm c}(L)/T_{\rm c}(\infty)-1 \propto AL^{-d/2}$, if $T_{\rm c}(L)$ is
defined as the temperature where $P_L(\phi)$ starts to develop a two-phase
structure. However, the next step in the development, due to Br\'ezin and
Zinn-Justin~\cite{bzj}, suggested that in the scaling limit $P_L(\phi) \propto
\exp(-\phi^4)$ at $T_{\rm c}$, since the shift of $T_{\rm c}$ as defined above
should only exhibit a scaling with a higher power of~$L^{-1}$, namely $T_{\rm
c}(L)/T_{\rm c}(\infty)-1 \propto L^{2-d}$, because it results from corrections
to the scaling description given in Eqs.\
(\ref{eq:scale2-m})--(\ref{eq:scale2-q}). According to Ref.~\cite{bzj}, the
asymptotic behavior is simply given by the homogeneous order parameter $M$ in
the finite system,
\begin{equation}
\label{eq:distr}
 P_L(M) = \exp \left[-L^d\left(\frac{1}{2}r_0 M^2 + uM^4\right)\right] \;,
\end{equation}
where $r_0=a_0 t$ changes sign at $T_{\rm c}$, $u$ is the dangerous irrelevant
variable mentioned above, both $a_0$ and $u$ being nonuniversal constants. From
Eq.~(\ref{eq:distr}) it is straightforward to derive that~\cite{bzj}
\begin{equation}
\label{eq:qvalue}
 \tilde{P}_Q(0) = 8\pi^2/\Gamma^4(1/4) \approx 0.456947 \;.
\end{equation}
However, the statements of CD imply that the continuum model considered
in~\cite{bzj} actually leads to a behavior differing from Eq.~(\ref{eq:distr})
and thus at this point also Eq.~(\ref{eq:qvalue}) seems without safe
foundation. CD obtain, in the large-$n$ limit of the $n$-vector model on the
lattice rather than in the continuum, a result for the scaling function
$P_\chi(x,y)$ [Eq.~(\ref{eq:scale-chi}); $x=t(L/\xi_0)^2$,
$y=(L/\ell_0)^{4-d}$] which is believed to be asymptotically exact, namely
\begin{equation}
\label{eq:pchi}
 P_\chi(x,y) = \frac{2}{J_0} 
   \left[ \delta(x,y) + \sqrt{[\delta(x,y)]^2+4y} \right]^{-1} \;,
\end{equation}
where $J_0$ is the interaction range of the $\phi^4$ model on the
$d$-dimensional hypercubic lattice (the lattice spacing being taken as unity
here throughout),
\begin{equation}
J_0 = \frac{1}{dL^d} \sum_{i,j} J_{ij} |\mathbf{r}_i-\mathbf{r}_j|^2 \;,
\end{equation}
and $\delta(x,y)$ is given by
\begin{equation}
\label{eq:delta}
 \delta(x,y) = x -y I_1(J_0^{-1} P_\chi^{-1}) \;,
\end{equation}
with the function $I_m(x)$, $m=1,2,\ldots,$ being
\begin{eqnarray}
\lefteqn{I_m(x) = 
 \frac{1}{(2\pi)^{2m}} \int_0^\infty dy\, y^{m-1} \exp(-xy/4\pi^2)} \nonumber\\
 && \times
 \left[\left(\frac{\pi}{y}\right)^{d/2} -
       \left(\sum_{p=-\infty}^{\infty} e^{-yp^2}\right)^d + 1 \right] \;.
\label{eq:int}
\end{eqnarray}
In terms of the Hamiltonian of the $n$-vector model with $n$-component vectors
$\phi_i$ on the lattice,
\begin{equation}
\label{eq:lathamil}
 H = \sum_i \left[ \frac{r_0}{2}\phi_i^2+u_0{(\phi_i^2)}^2 \right]
     + \sum_{i,j} \frac{J_{ij}}{2}(\phi_i-\phi_j)^2 \;,
\end{equation}
the characteristic lengths $\xi_0$, $\ell_0$ in Eqs.\
(\ref{eq:scale-m})--(\ref{eq:scale-q}) are given by
\begin{equation}
 \xi_0^2 = \frac{J_0}{a_0}(1+S_{\rm c}^{\rm b}), \quad 
 \ell_0^{d-4} = \frac{4u_0n}{J_0^2}\frac{1}{1+S_{\rm c}^{\rm b}} \;,
\end{equation}
with $r_0 = r_{0{\rm c}} + a_0 t$ and
\begin{equation}
 S_{\rm c}^{\rm b} = 
 u_0n \int d\mathbf{k} \, [\delta J(\mathbf{k})]^{-2}
 \;, \quad \delta J(\mathbf{k}) = J(\mathbf{0}) - J(\mathbf{k}) \;,
\label{eq:scb}
\end{equation}
where $J(\mathbf{k}) \equiv L^{-d} \sum_{i,j} J_{ij} \exp [-i\mathbf{k} \cdot
(\mathbf{r}_i-\mathbf{r}_j)]$.  For the $n$-vector model with $n=1$, which is
supposed to belong to the Ising universality class, comparable results are
obtained only to one-loop order in perturbation theory~\cite{cd_ijmpc}.
Although the results are not exact, their scaling structure is analogous to
Eqs.~(\ref{eq:pchi})--(\ref{eq:int}) and this structure is not expected to be
changed by the higher-order terms of the loop expansion. Defining reduced
moments
\begin{equation}
 \theta_m(Y) = \frac{\int_0^\infty d\phi \, \phi^m 
                     \exp\left[-\frac{1}{2}Y\phi^2 -\phi^4\right]}%
                    {\int_0^\infty d\phi \,
                     \exp\left[-\frac{1}{2}Y\phi^2 -\phi^4\right]}
\label{eq:theta}
\end{equation}
CD find~\cite{cd_ijmpc,cd_priv}
\begin{eqnarray}
\label{eq:pchi-ising}
 P_\chi(x,y) &=& \frac{1}{J_0}\frac{\theta_2(Y(x,y))}%
                                   {\sqrt{y+36I_2(\bar{r})y^2}}  \;,\\
 P_Q(x,y)    &=& \frac{[\theta_2(Y(x,y))]^2}{\theta_4(Y(x,y))}   \;,
\label{eq:scalefn-q}
\end{eqnarray}
with 
\begin{eqnarray}
 Y(x,y) &=& \left[x-12yI_1(\bar{r})
                 -144\theta_2(xy^{-1/2})I_2(\bar{r})y^{3/2}\right] /
 \nonumber \\
 && \left[y+36I_2(\bar{r})y^2\right]^{1/2} \;,
\label{eq:Y}
\end{eqnarray}
where $\bar{r} \equiv x + 12\theta_2(xy^{-1/2})y^{1/2}$. As should be clear
from what has been said above, the results (\ref{eq:pchi-ising})--(\ref{eq:Y})
should hold for sufficiently large~$L$.

Armed with these results we are now in a better position to reconsider the
question already posed in Ref.~\cite{bnpy}, namely to take the limit $y \to
0$. For this purpose we first consider the large-$n$ limit, where we can
rewrite Eq.~(\ref{eq:pchi}) as
\begin{equation}
\label{eq:p-largen}
 P_\chi(x,y) = \frac{1}{J_0\sqrt{y}} 
               \left[ \frac{\delta(x,y)}{2\sqrt{y}} +
                      \sqrt{1+[\delta(x,y)/(2\sqrt{y})]^2} \right]^{-1} \;.
\end{equation}
In the limit $y \to 0$ we see from Eq.~(\ref{eq:delta}) that
$P_\chi(x,y)$ depends, apart from the prefactor, on $x$ and $y$ through the
variable
\begin{eqnarray}
\lefteqn{\frac{\delta(x,y)}{2\sqrt{y}} \to \frac{x}{2\sqrt{y}} -
  \frac{1}{2}\sqrt{y}I_1(J_0^{-1}P_\chi^{-1})} \nonumber \\
  && = \frac{1}{2}tL^{d/2}\xi_0^{-2} \ell_0^{(4-d)/2} - 
       \frac{1}{2}(L/\ell_0)^{(4-d)/2} I_1(J_0^{-1}P_\chi^{-1}) \;. \nonumber\\
\label{eq:scalevar}
\end{eqnarray}
Thus we see that there exists a limit $t \to 0$, $L \to \infty$, $tL^{d/2}$
fixed, where the susceptibility scales exactly as postulated in
Eq.~(\ref{eq:scale2-chi}), since then the correction term of order
$(L/\ell_0)^{(4-d)/2}$ in Eq.~(\ref{eq:scalevar}) clearly is negligible
($J_0^{-1}P_\chi^{-1}$ tends toward zero in this limit, so the function $I_1$
approaches a finite constant).  Contrary to statements made by CD themselves,
\emph{viz.}\ that the structure of Eqs.\
(\ref{eq:scale2-m})--(\ref{eq:scale2-q}) is incorrect for the $\phi^4$ lattice
model, we rather think that they have proven(!) the correctness of
Eq.~(\ref{eq:distr}), in the limit specified above, at least for the large-$n$
limit, and gratifyingly there is no contradiction at all between Eqs.\
(\ref{eq:pchi})--(\ref{eq:scb}) and the ideas of Ref.~\cite{bnpy} that led to
Eqs.\ (\ref{eq:scale2-m})--(\ref{eq:scale2-q}). Of course, the strong merit of
the CD treatment is that it yields not only the scaling structure but also the
explicit scaling function and a full description of the corrections due to the
dangerous irrelevant variable~$u_0$.

We arrive at similar conclusions in the case $n=1$, though one must recall that
these results are only based on a one-loop order approximation. In the
considered limit $y \to 0$ the quantity $Y(x,y)$ in Eq.~(\ref{eq:Y}) reduces to
\begin{eqnarray}
Y(x,y) &\to& 
 \frac{x}{\sqrt{y}}\left[1 - 18 I_2(\bar{r})y \right] - 12I_1(\bar{r})\sqrt{y}
 \nonumber \\
 &=& tL^{d/2} \xi_0^{-2}\ell_0^{(4-d)/2}
      \left[1- 18 I_2(\bar{r}) (L/\ell_0)^{4-d} \right] \nonumber \\
 &&  \mbox{} - 12 I_1(\bar{r}) (L/\ell_0)^{(4-d)/2} \;,
\label{eq:scalevar2}
\end{eqnarray}
which is, apart from the additional $\mathcal{O}(L^{4-d})$ correction, fully
analogous to Eq.~(\ref{eq:scalevar}). In the limit of interest ($t \to 0$, $L
\to \infty$, $tL^{d/2}$ fixed), $\bar{r}$ vanishes and the functions $I_1$,
$I_2$ take finite values, so we see again that Eqs.\ (\ref{eq:scale2-chi})
and~(\ref{eq:scale2-q}) hold.  Moreover, one concludes that at the critical
temperature $Y(0,y \to 0) \to 0$ and hence also Eq.~(\ref{eq:qvalue}) holds, as
noted already by CD\@. It remains to be seen whether~(\ref{eq:qvalue}), which
is less general than the scaling structure of Eqs.\ 
(\ref{eq:scale2-m})--(\ref{eq:scale2-q}), holds to all orders in the loop
expansion.

In order to compare Eqs.\ (\ref{eq:theta})--(\ref{eq:Y}) to numerical Monte
Carlo data, it is clearly of interest to consider simple limiting cases of the
susceptibility, where then the nonuniversal parameters $\xi_0$ and $\ell_0$ can
be extracted. Since accurate Monte Carlo estimations of correlation lengths are
much more difficult to perform than estimations of the susceptibility we note
that in the large-$n$ limit [Eqs.\ (\ref{eq:pchi})--(\ref{eq:scb})] the bulk
susceptibility is~\cite{cd_physica}
\begin{equation}
 \chi_{\rm b} = -\frac{\partial^2 f_{\rm b}(t,h)}{\partial h^2}
                   = \frac{\xi_{\rm b}^2}{J_0}
                   = \frac{1+S_{\rm c}^{\rm b}}{a_0 t} \;,
\end{equation}
where $\xi_{\rm b}=\xi_0 t^{-\nu}$. The same result is obtained from Eqs.\
(\ref{eq:scale-chi}), (\ref{eq:pchi}) using that, at fixed small $t$,
$\delta(x,y)\approx x$ in the limit $L \to \infty$ and hence
\begin{equation}
 P_\chi(x,y \to 0) \approx (J_0 x)^{-1} \Rightarrow
 \chi = L^2/[J_0(L/\xi_{\rm b})^2] = \chi_{\rm b} \;.
\end{equation}
In contrast, at the critical temperature the result is
\begin{equation}
 J_0 \chi(T=T_{\rm c}) = L^{d/2} \ell_0^{(4-d)/2} \;.
\end{equation}
Thus, one can determine both parameters of interest, $\xi_0$ and $\ell_0$, from
the behavior of $\chi$ in two simple limits. The same procedure can also be
carried out in the $n=1$ case, considering the limit $y \to 0$ at fixed
small~$t$,
\begin{equation}
 \chi t = \frac{\xi_0^2}{J_0} \frac{x}{\sqrt{y+36I_2(\bar{r})y^2}} 
          \theta_2(Y(x,y)) 
        \stackrel{y \to 0}{\longrightarrow} \frac{\xi_0^2}{J_0} \;,
\label{eq:chi-xi0} 
\end{equation}
while in the finite-size scaling limit ($x=0$, $y$ small) one obtains for $d=5$
\begin{equation}
 \chi = \frac{L^2}{J_0} \frac{1}{\sqrt{y}} \theta_2(0)
      = \frac{L^{d/2}}{J_0\sqrt{\ell_0}} 
        \frac{\Gamma(\frac{3}{4})}{\Gamma(\frac{1}{4})} \;.
\label{eq:chi-l0}
\end{equation}
As we already noted, the interest of Eqs.\ (\ref{eq:theta})--(\ref{eq:Y}) is
not only the combination of the scaling structure, Eqs.\ 
(\ref{eq:scale2-m})--(\ref{eq:scale2-q}), in the limit $t \to 0$, $L \to
\infty$, $tL^{d/2} = {\rm const}$, but these equations also incorporate the
effect of the corrections to the lowest-mode approximation, which we would
recover if in Eq.~(\ref{eq:Y}) we had $Y(x,y)=xy^{-1/2}$.

\section{Comparison of the Chen--Dohm predictions with simulation results for
the \boldmath$d=5$ Ising model}
\label{sec:compare}
In Ref.~\cite{univ5d} only numerical data for the amplitude ratio~$Q$ have
been considered, with $5 \leq L \leq 22$.  In order to estimate the scaling
parameter $\ell_0$ we now analyze the corresponding data for the magnetic
susceptibility. Thus, we apply a finite-size expansion similar to Eq.~(3)
in~\cite{univ5d},
\begin{eqnarray}
\chi(T,L) &=& L^{d/2} \left( p_0 + p_1 \hat{t} L^{y_{\rm t}^*} +
                                    p_2 \hat{t}^2 L^{2y_{\rm t}^*}
                       \right. \nonumber \\
 && \phantom{L^{d/2} \left(\vphantom{L^{y^*}}\right.} 
 \left. \vphantom{L^{y^*}} + q_1 L^{y_{\rm i}} + q_2 L^{2y_{\rm i}} \right) \;,
\label{eq:chi-fit}
\end{eqnarray}
where $\hat{t}=t+\alpha L^{y_{\rm i}-y_{\rm t}}$ and $y_{\rm t}^*=y_{\rm t} -
y_{\rm i}/2$. So the term $\hat{t} L^{y_{\rm t}^*} = tL^{d/2} + \alpha
L^{(4-d)/2}$ just corresponds to the scaling variable in (\ref{eq:scalevar})
and~(\ref{eq:scalevar2}). The additional term in Eq.~(\ref{eq:scalevar2}) was
already mentioned in~\cite{univ5d} as the ``cross term'' $\hat{t}L^{y_{\rm t}^*
+ y_{\rm i}}$; in contrast to the analysis of~$Q$, it turns out to have a
negligibly small coefficient in the analysis of the susceptibility. The leading
power $L^{d/2}$ [Eq.~(\ref{eq:chi-l0})] has been confirmed numerically within
less than one percent in Ref.~\cite{thesis}. In our analysis we have kept this
power as well as the irrelevant exponent~$y_{\rm i}$ fixed. This yielded a
critical coupling $J/k_{\rm B}T_{\rm c} = 0.1139152~(4)$, in excellent
agreement with the value found in Ref.~\cite{univ5d} from keeping $Q$ fixed at
the zero-mode prediction~(\ref{eq:qvalue}), \emph{viz.}\ $J/k_{\rm B}T_{\rm c}
= 0.1139150~(4)$ (numbers in parentheses denote the uncertainty in the last
decimal places). Furthermore, we found $y_{\rm t}^*=2.53~(4)$, very close
to~$d/2$, and $p_0 = 1.86~(7)$. The quality of the fit in terms of the $\chi^2$
criterion was $\chi^2/{\rm DOF} = 1.06$.  In order to improve the accuracy of
our estimate for~$p_0$, we have repeated the analysis with $y_{\rm t}^*$ fixed
at~$d/2$, finding $J/k_{\rm B}T_{\rm c} = 0.1139155~(2)$ and $p_0=1.91~(2)$
($\chi^2/{\rm DOF} = 1.05$). All analyses were obtained with $5 \leq L \leq
22$; upon omitting the smallest system sizes, a very similar estimate for $p_0$
was obtained, with a minor increase in the uncertainty.  For a more detailed
analysis we refer to~\cite{5dpaper,thesis}. From our estimate for~$p_0$ and
Eq.~(\ref{eq:chi-l0}) we find, using $J_0=2 J/k_{\rm B}T$, $\ell_0=0.603~(13)$.
For the sake of clarity, it is stressed that this estimate for $\ell_0$ thus
pertains to the thermodynamic limit and is \emph{not} a finite-size quantity.

It is also possible to extract $\xi_0$ from the Monte Carlo data. However, here
we use the series-expansion result from Ref.~\cite{guttmann} for this
purpose. Assuming the mean-field exponent $\gamma=1$, it was found that
asymptotically $\chi = A/(1 - v/v_{\rm c})$ with $A=1.311~(9)$ and
$v=\tanh(J/k_{\rm B}T)$. Rewriting this in terms of the reduced
temperature~$t$, we have $\chi = A [ \cosh(J/k_{\rm B}T_{\rm c})\sinh(J/k_{\rm
B}T_{\rm c}) / (J/k_{\rm B}T_{\rm c}) ] t^{-1} = 1.322 t^{-1}$ and
Eq.~(\ref{eq:chi-xi0}) shows that $\xi_0=0.549~(2)$.

\begin{figure}
\leavevmode \centering \epsfxsize \columnwidth
\epsfbox{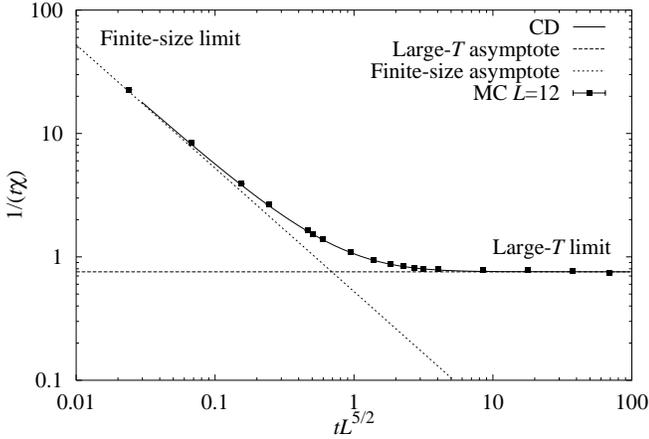}
\caption{Log--log plot of $1/(t \chi)$ versus $tL^{5/2}$, for the
five-dimensional nearest-neighbor Ising model. Squares denote Monte Carlo data
as mentioned in the text, dashed lines represent Eqs.\ (\ref{eq:chi-xi0})
and~(\ref{eq:chi-l0}), respectively, with the parameters from
Eq.~(\ref{eq:l-xi}). The finite-size asymptote included here refers to the
limit $L\to\infty$ and has been estimated as described after
Eq.~(\protect\ref{eq:chi-fit}).}
\label{fig:chi-param}
\end{figure}

Figure~\ref{fig:chi-param} now shows the log--log plot of $(t\chi)^{-1}$ versus
the scaling variable $tL^{d/2}$, using data for $T \geq T_{\rm c}$ only (in
view of the very accurate estimates of the critical coupling, the errors due to
the inaccuracy of $T_{\rm c}$ are not of major concern here). Available data
for smaller system sizes have been omitted from this graph, because the rather
strong deviations from scaling noted already in Ref.~\cite{binder5d} would
obscure its main purpose, namely to illustrate the use of the limits
(\ref{eq:chi-xi0}), (\ref{eq:chi-l0}) to extract $\xi_0$ and $\ell_0$.  Of
course, due to the corrections to scaling included in Eq.~(\ref{eq:chi-fit})
the Monte Carlo data for $L=12$ should not converge to the finite-size
asymptote for $L\to\infty$, but to a slightly shifted straight line. However,
on the scale of Fig.~\ref{fig:chi-param} the finite-size asymptotes for $L=12$
and $L\to\infty$ are indistinguishable. Because of their central interest we
repeat our estimates
\begin{equation}
\label{eq:l-xi}
 \ell_0 = 0.603~(13) \;, \quad \xi_0 = 0.549~(2) \;.
\end{equation}
The amplitude $\xi_0^2 \sqrt{\ell_0}$ which normalizes the scaling variable
$tL^{d/2}$ [cf.\ Eq.~(\ref{eq:scalevar2})] becomes 0.234~(4).

In the following graphs, also Monte Carlo data from Ref.~\cite{binder5d}
($L=4$) and Ref.~\cite{rickwardt} ($L=8,12$) are included and it was found that
all the Monte Carlo data are, within their statistical errors, nicely
compatible with each other.  We have omitted the data of Ref.~\cite{rickwardt}
for $L \geq 13$ here, since the rather irregular behavior found for the
specific heat and the cumulant intersections for these system sizes indicates
that these data suffer from statistical inaccuracies due to critical slowing
down. Note that Ref.~\cite{rickwardt} used a single-spin-flip Metropolis
algorithm, whereas in Refs.~\cite{univ5d,5dpaper,thesis} a single-cluster
algorithm was applied. Available data from Refs.~\cite{mon,parisi} are
restricted to temperatures very close to $T=T_{\rm c}$ and hence are unsuitable
for our purposes.

\begin{figure}
\leavevmode \centering \epsfxsize \columnwidth
\epsfbox{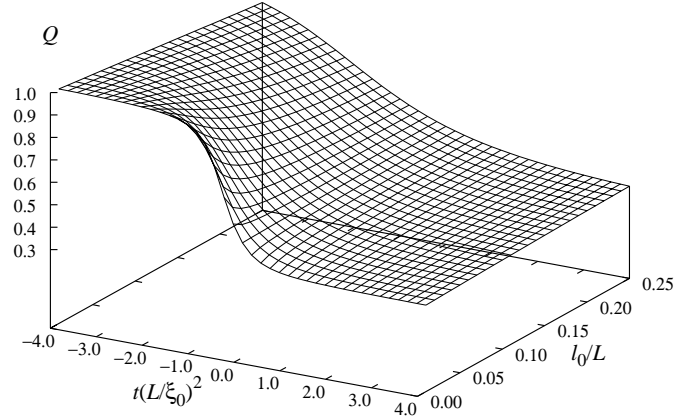}
\caption{Plot of the moment ratio $Q \equiv \langle M^2\rangle^2 / \langle M^4
\rangle$ for $d=5$ as a function of the two variables $x=tL^2/\xi_0^2$ and
$y=(\ell_0/L)^{d-4}$, according to Eqs.\ (\ref{eq:theta}), (\ref{eq:scalefn-q})
and~(\ref{eq:Y}).}
\label{fig:q-3d}
\end{figure}

We now focus on the quantity~$Q$, Eq.~(\ref{eq:scale-q}), since the scaling
behavior of this quantity has been so controversial. Figure~\ref{fig:q-3d}
gives a plot of the CD function~(\ref{eq:scalefn-q}), keeping both $x$ and $y$
as separate variables. One can see that for $x$ large and negative $Q=1$ as it
must be and for $x$ large and positive $Q=1/3$, irrespective of $y$. This
simply reflects the trivial properties of the low- and high-temperature phases,
respectively. For $|t(L/\xi_0)^2| < 1$, however, a clear $y$ dependence is
seen.

\begin{figure}
\leavevmode \centering \epsfxsize \columnwidth
\epsfbox{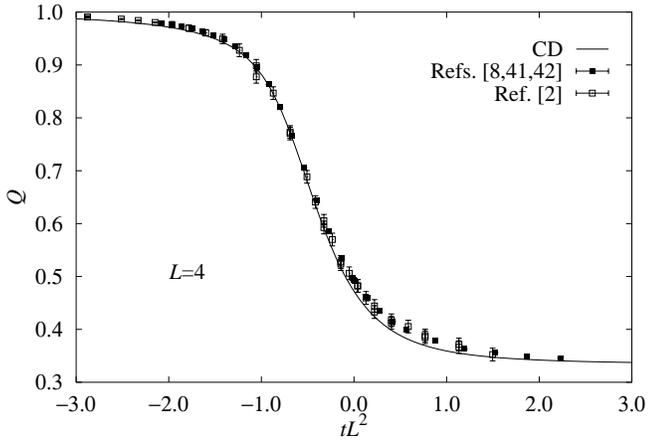} \\ (a) \\ 
\leavevmode \centering \epsfxsize \columnwidth
\epsfbox{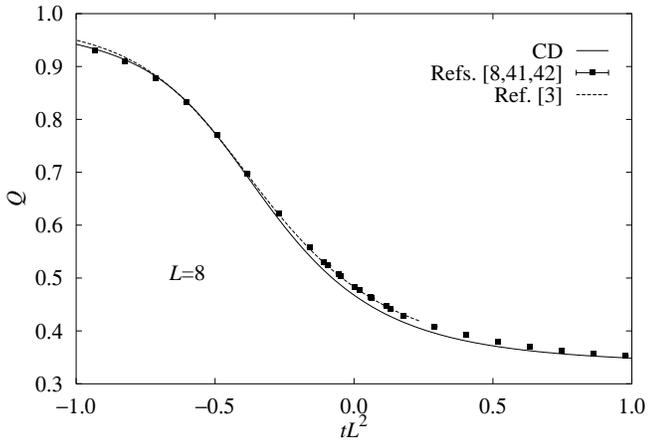} \\ (b) \\
\leavevmode \centering \epsfxsize \columnwidth
\epsfbox{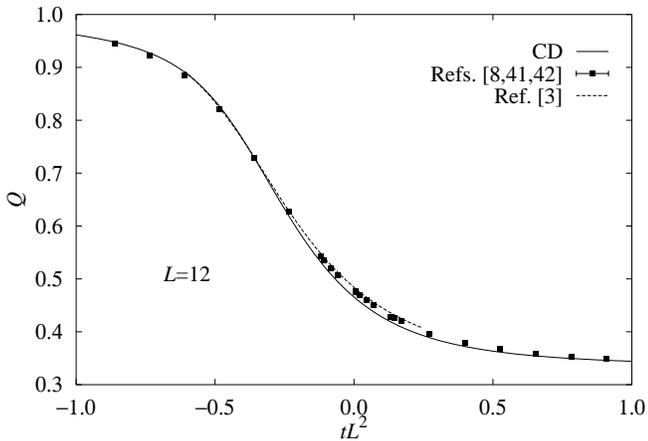} \\ (c) \\
\caption{Plot of $Q$ versus $tL^2$ for (a) $L=4$, (b) $L=8$ and (c) $L=12$. The
full curves denote the predictions of Ref.~\cite{cd_ijmpc}. Monte Carlo data
generated at specific temperatures, taken from Refs.\ \cite{binder5d} and
\cite{univ5d,5dpaper,thesis} are shown as open or full squares, respectively,
while the histogram extrapolation data of Ref.~\cite{rickwardt} are shown as a
broken curve.}
\label{fig:q-l-scale}
\end{figure}

\begin{figure}
\leavevmode \centering \epsfxsize \columnwidth
\epsfbox{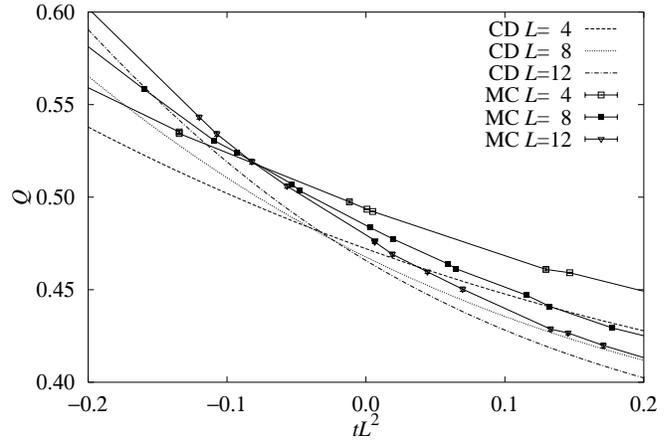}
\caption{Magnified plot of $Q$ versus $tL^2$ near $tL^2=0$, cf.\
Fig.~\ref{fig:q-l-scale}, to demonstrate the occurrence of spurious cumulant
intersections.}
\label{fig:q-intersect}
\end{figure}

In Fig.~\ref{fig:q-l-scale} we compare the various Monte Carlo data to the CD
function for $Q$ as a function of $tL^2$ (i.e., the variable proportional to
$x$). Note that in these plot there are \emph{no adjustable parameters}
whatsoever, so the agreement is at first sight very remarkable. At second
sight, however, one does observe that there are slight but systematic
deviations between theory and simulation, which have consequences for the
intersection of the amplitude ratios for different system sizes.
Figure~\ref{fig:q-intersect} demonstrates that both the CD function and the
Monte Carlo data behave qualitatively similar: for a range of sizes ($4 \leq L
\leq 12$) there is almost a common intersection point, but it occurs at a
\emph{negative} value of $tL^2$ and consequently the corresponding ordinate
value~$Q_{\rm int}$ is significantly larger than the predicted asymptotic
value~(\ref{eq:qvalue}). While this spurious value of the CD function, for the
range of system sizes considered here, is about $Q_{\rm int} \approx 0.48$, it
lies around $Q_{\rm int} \approx 0.52$ for the Monte Carlo data; Rickwardt
\emph{et al.}~\cite{rickwardt} quoted $Q_{\rm int} \approx 0.49~(1)$, including
data up to $L=17$. The lesson to be learned from this graph is threefold:
(i)~One must not pay too much attention to the value of such a cumulant
intersection if one does not have a sufficiently large range of linear
dimensions at one's disposal. (ii)~The CD function is a nice analytical example
of a function that does produce a spurious ``intersection'', as pointed out
already in Ref.~\cite{cd_ijmpc}: Although it looks so convincing on the graph,
one knows that in the asymptotic limit the intersection occurs at $t=0$ and
yields $Q \approx 0.457$ [Eq.~(\ref{eq:qvalue})].  (iii)~The CD function
produces the same trend as the Monte Carlo data only qualitatively, but not
quantitatively.

\begin{figure}
\leavevmode \centering \epsfxsize \columnwidth
\epsfbox{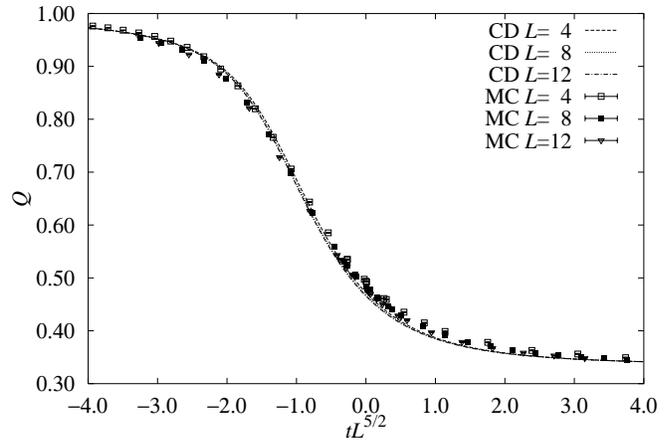}
\caption{Plot of $Q$ as a function of $tL^{5/2}$, including both Monte Carlo
data for $L=4,8,12$ and the CD functions. Note that the zero-mode curve,
resulting from setting $Y=xy^{-1/2}$ in Eqs.\ (\ref{eq:theta}),
(\ref{eq:scalefn-q}) and~(\ref{eq:Y}), practically coincides with the CD curve
for $L=12$ already, since $\ell_0$ [Eq.~(\ref{eq:l-xi})] is so small.}
\label{fig:q-scale}
\end{figure}

What is the consequence of these results for the asymptotic scaling,
Eq.~(\ref{eq:scale2-q})?  This question is addressed in Fig.~\ref{fig:q-scale},
where the data from Fig.~\ref{fig:q-l-scale} are replotted as a function of the
scaling variable $tL^{d/2}$: It is seen that the data for $L=4$ deviate from
scaling in a systematic way, while for $L=8,12$ the data scale already rather
nicely, although they are still a little bit set off in comparison to the
theoretical scaling curves. Note that on these large scales one cannot
distinguish the CD curve for $L=12$ from the lowest-mode result! The general
trend appears that in the neighborhood of $T=T_{\rm c}$
Eq.~(\ref{eq:scalefn-q}) yields a too small value for $Q(L)$.

\begin{figure}
\leavevmode \centering \epsfxsize \columnwidth
\epsfbox{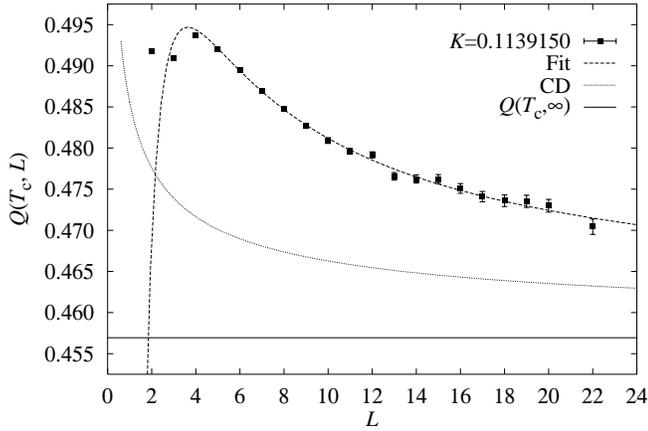}
\caption{Plot of $Q(T_{\rm c},L)$ versus $L$. Data points are the Monte Carlo
results from Ref.~\cite{univ5d}, including statistical errors. The broken curve
is the empirical fit, as described in Ref.~\cite{univ5d}, while the horizontal
line denotes the asymptotic result of Eq.~(\ref{eq:qvalue}). The CD prediction
is shown as a dotted line.}
\label{fig:q-kc}
\end{figure}

In order to highlight the differences, we now amend the plot of $Q$ at $T_{\rm
c}$ as a function of $L$, which was shown in Ref.~\cite{univ5d}, by the
prediction that would follow from CD (see Fig.~\ref{fig:q-kc}): One indeed
observes that the result of CD underestimates the differences between $Q(T_{\rm
c},L)$ and $Q(T_{\rm c},\infty)$ distinctly---it basically yields a
$1/\sqrt{L}$ correction, while the additional $1/L$ term resulting from CD is
very small, unlike the rather pronounced $1/L$ correction that was found in the
the empirical fit of Ref.~\cite{univ5d}. Another, more tentative, way to
quantify the differences is by adjusting $\ell_0$ such that the CD curve yields
a reasonable description of the numerical data. It turns out that a value as
high as $\ell_0 \approx 3.2$ (instead of the estimated value $0.603 \pm 0.013$)
is required to find some agreement in the region $12 \leq L \leq 22$. It
clearly must be waited for a loop expansion to second order---which will yield
additional $1/L$ corrections of so far unknown magnitude---before one can draw
final conclusions about the agreement between theory and simulation (or lack
thereof).

\begin{figure}
\leavevmode \centering \epsfxsize \columnwidth
\epsfbox{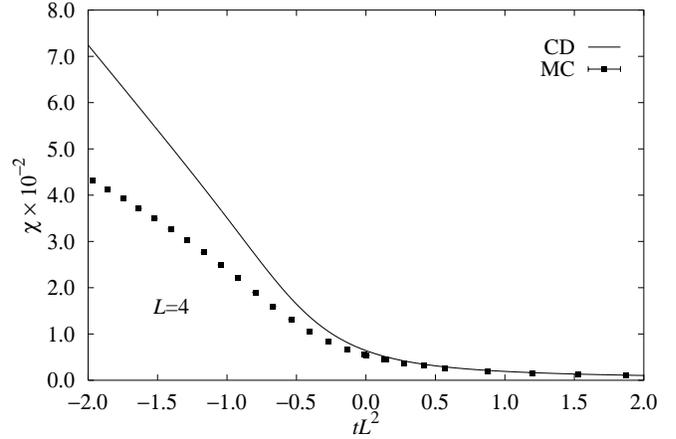} \\ (a) \\ 
\leavevmode \centering \epsfxsize \columnwidth
\epsfbox{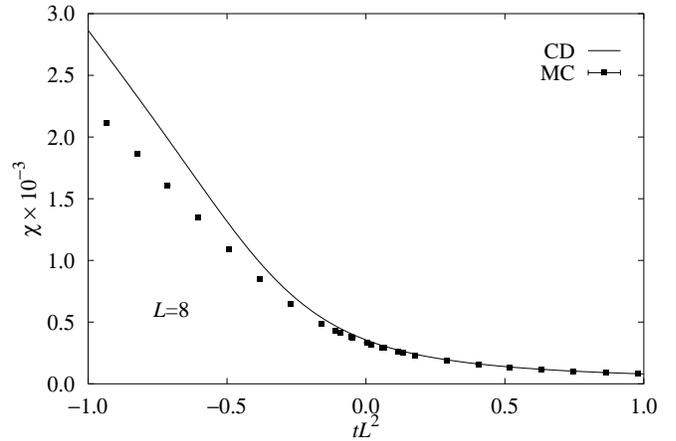} \\ (b) \\
\leavevmode \centering \epsfxsize \columnwidth
\epsfbox{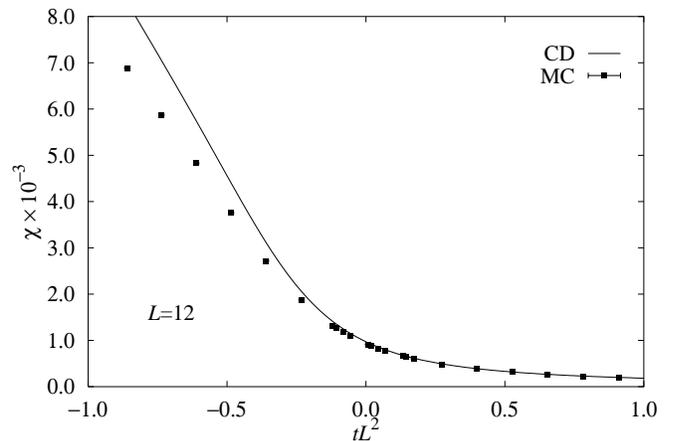} \\ (c) \\
\caption{Plot of $\chi$ versus $tL^2$ for (a) $L=4$, (b) $L=8$ and (c)
$L=12$. The full curves denote the predictions of Ref.~\cite{cd_ijmpc},
Eq.~(\ref{eq:pchi-ising}), while the squares are Monte Carlo data of
Refs.~\cite{univ5d,5dpaper,thesis}.}
\label{fig:chi}
\end{figure}

While in the comparison of the temperature dependence of $Q$ as predicted by CD
theory to the simulation results we have seen most of the disagreement for
$t>0$, Fig.~\ref{fig:chi} shows that much more drastic deviations between
theory and simulation occur for $\chi$ in the regime $t<0$. The fact that for
$t>0$ there is perfect agreement for $L=12$ is no surprise, of course, since
these data have been of central relevance for the fit in
Fig.~\ref{fig:chi-param} that yielded $\xi_0$ and $\ell_0$. It is clear that
perhaps a better overall fit of the data is reached if one would fit $\ell_0$
to describe the behavior of $\chi$ for $tL^2$ large and negative, but then the
behavior for $t > 0$ would deteriorate. Let us note in passing that the term
``susceptibility'' is just used for convenience here: Below~$T_{\rm c}$ the
real (reduced) susceptibility is of course given by $L^d (\langle M^2 \rangle -
\langle |M|^2) \rangle$.

\begin{figure}
\leavevmode \centering \epsfxsize \columnwidth
\epsfbox{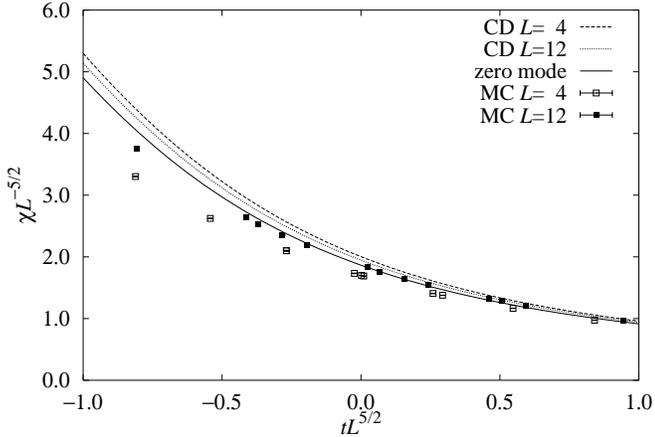}
\caption{Plot of $\chi L^{-5/2}$ versus $tL^{5/2}$ including Monte Carlo data
for $L=4$ and~$L=12$. Broken curves shown corresponding CD predictions, full
curve is the zero-mode result.}
\label{fig:chi-l-scale}
\end{figure}

\begin{figure}
\leavevmode \centering \epsfxsize \columnwidth
\epsfbox{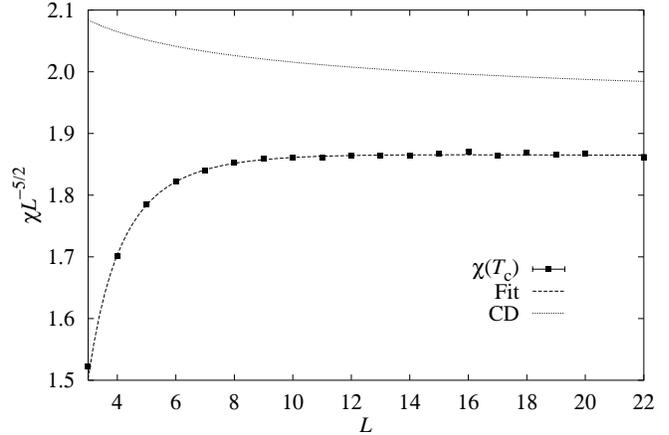}
\caption{Plot of $\chi(T_{\rm c}) L^{-5/2}$ versus $L$ (where $J/k_{\rm
B}T_{\rm c}=0.1139150$). The dashed curve is a fit to Eq.~(\ref{eq:chi-fit}) as
described in the text.  The dotted curve is the CD
result~(\ref{eq:pchi-ising}).}
\label{fig:chi-kc}
\end{figure}

Figure~\ref{fig:chi-l-scale} shows then a plot of $\chi L^{-5/2}$ versus
$tL^{5/2}$, comparing Monte Carlo data for $L=4$ and $L=12$~\cite{univ5d} with
corresponding predictions of the CD theory and the ``zero-mode'' curve. Again
systematic deviations between CD theory and simulations are apparent: while the
theory~\cite{cd_ijmpc} converges to the zero-mode result from above, the Monte
Carlo results fall clearly \emph{below} the zero-mode result and nearly
coincide with it for $L=12$. This coincidence can be understood from a closer
consideration of $\chi(T_{\rm c})L^{-5/2}$ versus $L$ (Fig.~\ref{fig:chi-kc}):
After a rapid increase from below to a value already close to the asymptotic
value, this quantity flattens around $L=12$ and then slowly approaches (not
necessarily in a monotonic way) its limiting value.  In the whole region shown,
the CD curve \emph{qualitatively disagrees} with the data---this disagreement
clearly cannot be remedied by a different adjustment of the parameters, because
a monotonic decrease (close to a $1/\sqrt{L}$ behavior) is an intrinsic feature
of Eqs.\ (\ref{eq:theta})--(\ref{eq:Y}) and also occurs in the large-$n$ limit
[Eq.~(\ref{eq:p-largen})]. The deviation at $L=22$ cannot be explained from a
mis-adjustment of $\ell_0$, since both curves approach the same limiting value
for $L \to \infty$, where all finite-size corrections must vanish.  Thus, if
$\ell_0$ would have been chosen such that the CD curve coincides with the Monte
Carlo result for $L=22$, a mismatch would have occurred at $L \to \infty$,
which is clearly impossible.

Of course, discrepancies between finite-size data for very small linear
dimensions (such as $L=4$ and $L=8$) and the CD theory
[Eqs.~(\ref{eq:pchi-ising})--(\ref{eq:Y})], which only fully captures the
leading zero-mode result and the first correction terms (of order $L^{-1/2}$)
to it, would not be an argument against the usefulness of the theory. However,
Figs.~\ref{fig:q-kc} and~\ref{fig:chi-kc} clearly reveal that even for $L=22$
one is still far from the regime where the CD theory satisfactorily describes
the MC data.

%\newpage

\section{Concluding remarks}
\label{sec:concl}
In this paper Monte Carlo results for five-dimensional Ising lattices have been
reanalyzed and compared to recent theoretical predictions obtained by Chen and
Dohm, in an attempt to clarify a somewhat controversial discussion. Our results
can be summarized as follows:

(i) The CD theory reduces in the limit $t \to 0$, $L \to \infty$, $tL^{d/2}$
fixed, to the scaling structure proposed originally by Binder \emph{et
al.}~\cite{bnpy} and explicitly illustrates the mechanism of multiplicative
``renormalization'' of variables by a dangerous irrelevant variable. In
addition, it yields both the asymptotic scaling functions of various moments of
the order parameter distribution as a function of the variable $tL^{d/2}$ and
the leading corrections to it, which are of order $L^{(4-d)/2}$. However, the
comparison with the Monte Carlo data indicates that strong subleading
corrections (of order $1/L$ for $d=5$) are present as well, which are not
predicted by the CD theory, and one would need much larger~$L$ than accessible
here ($L_{\rm max}=22$) in order that these subleading corrections are visible.
So a quantitative agreement between theory and simulation is still far out of
reach!

(ii) The question must be asked to what extent the $\phi^4$ model on a lattice
for $d>4$ yields the same behavior as the Ising model. Of course, one can take
the parameters $r_0 \to -\infty$, $u_0 \to \infty$ in Eq.~(\ref{eq:lathamil})
in such a proportion that the model precisely reduces to the Ising model (for a
discussion see, e.g., Ref.~\cite{milchev}). At this point, we have not
attempted to deal with this problem.

(iii) The CD theory yields a nice illustrative example how one can be misled by
apparent cumulant intersections which converge to the exact result extremely
slowly as $L \to \infty$. It is rather likely that this is the reason for the
difficulties noted in the Monte Carlo studies in
Refs.~\cite{bnpy,binder5d,rickwardt}. Both the CD theory and the simulations
give clear evidence that for a full understanding of the problem a variation of
parameters over a broad range is desirable, including the behavior both above
and below $T_{\rm c}$, as well as at $T_{\rm c}$. Corrections to the leading
scaling behavior need careful consideration, which was already one of the
central messages of Refs.~\cite{renorm,5dcomm,univ5d}.

\subsection*{Acknowledgement}
Stimulating discussions with Profs.\ V. Dohm and X.S. Chen are gratefully
acknowledged, as well as information on their work prior to publication.

%\begin{figure*}
%\caption{}
%\label{}
%\end{figure*}

%\begin{table}
%\caption{}
%\label{}
%\begin{tabular}{lll}
%\hline\noalign{\smallskip}
%\end{tabular}
%\end{table}


\begin{thebibliography}{99}
\bibitem{bnpy} K. Binder, M. Nauenberg, V. Privman, A.P. Young, Phys. Rev. B
  {\bf 31}, 1498 (1985).
\bibitem{binder5d} K. Binder, Z. Phys. B {\bf 61}, 13 (1985).
\bibitem{rickwardt} Ch. Rickwardt, P. Nielaba, K. Binder,
  Ann. Phys. (Leip\-zig) {\bf 3}, 483 (1994).
\bibitem{mon} K.K. Mon, Europhys. Lett. {\bf 34}, 399 (1996).
\bibitem{parisi} G. Parisi, J.J. Ruiz-Lorenzo, Phys. Rev. B {\bf 54}, R3698
  (1996); {\bf 55}, 6082(E) (1997).
\bibitem{5dcomm} E. Luijten, Europhys. Lett. {\bf 37}, 489 (1997).
\bibitem{5drepl} K.K. Mon, Europhys. Lett. {\bf 37}, 493 (1997).
\bibitem{univ5d} H.W.J. Bl\"ote, E. Luijten, Europhys. Lett. {\bf 38}, 565
  (1997).
\bibitem{fisher-fss} M.E. Fisher, in {\it Critical Phenomena}, edited by
  M.S. Green (Academic, N.Y., 1971).
\bibitem{fisher_barber} M.E. Fisher, M.N. Barber, Phys. Rev. Lett. {\bf 28},
  1516 (1972).
\bibitem{binder72} K. Binder, Physica {\bf 62}, 508 (1972).
\bibitem{binder-cum} K. Binder, Z. Phys. B {\bf 43}, 119 (1981).
\bibitem{brezin} E. Br\'ezin, J. Phys. (Paris) {\bf 43}, 15 (1982).
\bibitem{privfish} V. Privman, M.E. Fisher, J. Stat. Phys. {\bf 33}, 385
  (1983).
\bibitem{botet} R. Botet, R. Jullien, P. Pfeuty, Phys. Rev. Lett. {\bf 49}, 478
  (1982); R. Botet, R. Jullien, Phys. Rev. B {\bf 28}, 3955 (1983).
\bibitem{barber} M.N. Barber, in \textit{Phase Transitions and Critical
  Phenomena}, Vol. 8, edited by C. Domb and J.L. Lebowitz (Academic, London,
  1983).
\bibitem{bzj} E. Br\'ezin, J. Zinn-Justin, Nucl. Phys. B {\bf 257} [FS14],
  867 (1985).
\bibitem{rudnick} J. Rudnick, H. Guo, D. Jasnow, J. Stat. Phys. {\bf 41},
  353 (1985).
\bibitem{rudnick2} J. Rudnick, G. Gaspari, V. Privman, Phys. Rev. B {\bf 32},
  7594 (1985); J. Shapiro, J. Rudnick, J. Stat. Phys. {\bf 43}, 51 (1986).
\bibitem{huhn} W. Huhn, V. Dohm, Phys. Rev. Lett. {\bf 61}, 1368 (1988).
\bibitem{binder_wang} K. Binder, J.S. Wang, J. Stat. Phys. {\bf 55}, 87 (1989).
\bibitem{privman_book} V. Privman (ed.), \textit{Finite Size Scaling and
  Numerical Simulation of Statistical Systems} (World Scientific, Singapore,
  1990).
\bibitem{binder_ft} K. Binder, in \textit{Computational Methods in Field
  Theory}, edited by H. Gausterer and C.B. Lang (Springer, Berlin, 1992).
\bibitem{dohm93} V. Dohm, Physica Scripta {\bf T49}, 46 (1993).
\bibitem{mon_binder} K.~K. Mon, K. Binder, Phys.\ Rev.\ E {\bf 48}, 2498
  (1993).
\bibitem{dohm} S. Dasgupta, D. Stauffer, V. Dohm, Physica A {\bf 213}, 368
  (1995); A. Esser, V. Dohm, X.S. Chen, Physica A {\bf 222}, 355 (1995);
  X.S. Chen, V. Dohm, A. Talapov, Physica A {\bf 232}, 375 (1996); X.S. Chen,
  V. Dohm, N. Schultka, Phys. Rev. Lett. {\bf 77}, 3641 (1996); X.S. Chen,
  V. Dohm, Physica A {\bf 235}, 555 (1997); X.S. Chen, V. Dohm,
  Int. J. Mod. Phys. B {\bf 12}, 1277 (1998).
\bibitem{renorm} E. Luijten, H.W.J. Bl\"ote, Phys. Rev. Lett. {\bf 76},
  1557 (1996); {\bf 76}, 3662(E) (1996).
\bibitem{medran} E. Luijten, H.~W.~J. Bl\"ote, K. Binder, Phys.\ Rev.\ E
  {\bf 54}, 4626 (1996); {\bf 56}, 6540 (1997).
\bibitem{cd_physica} X.S. Chen, V. Dohm, Physica A {\bf 251}, 439 (1998).
\bibitem{cd_epjb} X.S. Chen, V. Dohm, Eur. Phys. J. B (to be published).
\bibitem{cd_ijmpc} X.S. Chen, V. Dohm, preprint ``Finite-size effects in the
  $\phi^4$ field and lattice theory above the upper critical dimension'', to be
  published in Int. J. Mod. Phys. C.
\bibitem{univfss} E. Luijten, preprint ``Test of renormalization predictions
  for universal finite-size scaling functions''.
\bibitem{binder_ciccotti} K. Binder, G. Ciccotti (eds.), \textit{Monte Carlo
  and Molecular Dynamics of Condensed Matter Systems} (Societ\`a Italiana di
  Fisica, Bologna, 1996).
\bibitem{binder_rpp} K. Binder, Rep. Prog. Phys. {\bf 60}, 487 (1997).
\bibitem{gausterer} H. Gausterer, C.B. Lang (eds.), \textit{Computational
  Methods in Field Theory} (Springer, Berlin, 1992).
\bibitem{meyer} H. Meyer-Ortmanns, Rev. Mod. Phys. {\bf 68}, 473 (1996).
\bibitem{fisher_rmp} M.E. Fisher, Rev. Mod. Phys. {\bf 46}, 597 (1974);
  {\bf 70}, 653 (1998).
\bibitem{qft} J. Zinn-Justin, {\it Quantum Field Theory and Critical
  Phenomena}, 3rd edn. (Clarendon, Oxford, 1996).
\bibitem{wegner_riedel} F.J. Wegner, E.K. Riedel, Phys. Rev. B {\bf 7}, 248
  (1973).
\bibitem{deutsch} H.P. Deutsch, K. Binder, Macromolecules {\bf 25}, 6214
  (1992); J. Phys. II (France) {\bf 3}, 1049 (1993).
\bibitem{5dpaper} E. Luijten, H.W.J. Bl\"ote, to be published.
\bibitem{thesis} E. Luijten, \textit{Interaction Range, Universality and the
  Upper Critical Dimension} (Delft University Press, Delft, 1997).
\bibitem{kvlh} H.J.F. Knops, J.M.J. van Leeuwen, P.C. Hemmer,
  J. Stat. Phys. {\bf 17}, 197 (1977).
\bibitem{fisher_review} M.E. Fisher, in \textit{Critical Phenomena}, edited by
  F.J.W. Hahne (Springer, Berlin, 1983).
\bibitem{cd_priv} X.S Chen, V. Dohm, private communication.
\bibitem{guttmann} A.J. Guttmann, J. Phys. A {\bf 14}, 233 (1981).
\bibitem{milchev} A. Milchev, D.W. Heermann, K. Binder, J. Stat. Phys. {\bf
  44}, 749 (1986).
\end{thebibliography}
\end{document}